\documentclass[amsmath,superscriptaddress,showpacs,prl,twocolumn] {revtex4}
\usepackage{bm}
\usepackage{enumerate}
\usepackage{graphicx}
\usepackage[dvips]{epsfig}
\usepackage{epsf}
\usepackage{amsmath}

\makeatletter
\newcommand{\Rmnum}[1]{\expandafter\@slowromancap\romannumeral #1@}
\makeatletter

\begin{document}
\title{Landau levels for electromagnetic wave}
\author{Vladimir A. Zyuzin}
\affiliation{Department of Physics and Astronomy and Nebraska Center for Materials and Nanoscience,
 University of Nebraska, Lincoln, Nebraska 68588, USA}
\affiliation{Department of Physics and Astronomy, Texas A$\mathrm{\&}$M University, College Station, Texas 77843-4242, USA}
\begin{abstract}
In this paper we show that the frequencies of propagating electromagnetic wave (photon) in rotating dielectric medium obey Landau quantization. 
We show that the degeneracy of right and left helicities of photons is broken on the lowest Landau level. 
In homogeneous space this level is shown to be helical, i.e. left and right helical photons counter-propagate. 
This leads to helical vortical effect for photons.
\end{abstract}
\maketitle

\textit{\underline{Introduction}}. 
Photons are spin- 1 massless particles and are described by helicity, a scalar product of spin and propagtion direction, which can only take $+1$ or $-1$ values. It is a natural spin-momentum locking of a photon. 
Therefore, when photon's momentum direction is adiabatically changed and returned to original direction, the spin acquires a phase - Berry phase \cite{Berry}. 
This phase was experimentally observed in a system of coiled optical fiber \cite{ChiaoWu, TomitaChiao}. 
Other photon properties due to Berry phase of photon were studied in \cite{Bliokh04, Bliokh08, Stone}.
Due to non-trivial Berry phases of photons there exist chiral photon edge states  \cite{QHE, exp1, exp2} which are analogous to quantum Hall systems. Also an analog of topological insulator for photons can be drawn, 
see a problem to $\S 68$ of \cite{LL8}, \cite{Agranovich_Ginzburg}, and \cite{Bliokh_Smirnova_Nori}.

Moreover, it is understood that the motion of the medium is an effective vector potential seen by the photons propagating in the medium, see $\S 57$ of \cite{LL8}, and \cite{LP}. 
Using this concept, the Aharonov-Bohm effect for photons was theoretically proposed in \cite{AB}. 
In this paper we explicitly show that the equation describing propagation of photons inside a uniformly rotating dielectric medium has a solution similar to the solution of the Schr\"{o}dinger equation for electron in magnetic field, the Landau wave-functions and corresponding Landau levels \cite{LL3}.
We find gapless helical zeroth Landau level.
In which case, photons with opposite helicities counter-propagate and do not mix with each other. 
Due to that there is a finite temperature non-zero helicity current in a gas of photons, the so-called helical vortical effect (or chiral vortical effect).
The helical Landau level is strickingly analogous to the chiral zeroth Landau level of three dimensional Dirac fermions in magnetic field, for example see \cite{NN}. 
This helical Landau level is called zero mode whose existence is due to non-trivial topology (Berry phase) of a photon.
Helical vortical effect for photons is analogous to chiral magnetic effect for Dirac fermions, see \cite{Kharzeev} for a review.

In a recent experimental work creation of synthetic Landau levels for photons in optical resonators was reported \cite{synthetic1}.
Another proposal of magnetic fluid of photons based on non-linear effects has recently been put forward, see \cite{synthetic2}.
Present paper proposes a simple picture of creating Landau levels for photons in rotating medium and hopefully it will inspire further research. It is possible that the Landau level for photons and the helical vortical effect can be observed in experiments with slow light \cite{slowlight}.

\textit{\underline{Landau levels for electromagnetic waves}}.
Maxwell equations describing propagation of electro-magnetic wave in the dielectric medium described by constant $\epsilon$ and $\mu$ in the absence of currents and charges are
\begin{align}
&{\bm\nabla}\cdot{\bf B} = 0, ~~ {\bm\nabla}\times{\bf H} = \frac{1}{c} \frac{\partial {\bf D}}{\partial t}, \\
&{\bm\nabla}\cdot{\bf D} = 0, ~~ {\bm\nabla}\times{\bf E} = -\frac{1}{c} \frac{\partial {\bf B}}{\partial t}.
\end{align}
Assume that the dielectric medium is moving with a speed ${\bf v}$. 
In the limit $\vert {\bf v}\vert/c \ll 1$, to the lowest order in $\vert {\bf v}\vert/c$, see  $\S 57$  of \cite{LL8}, we write
\begin{align}
& {\bf D} = \epsilon {\bf E}  + \frac{\epsilon\mu - 1}{c}\left[{\bf v}\times {\bf H} \right],
\\
& {\bf B} = \mu {\bf H}  - \frac{\epsilon\mu - 1}{c}\left[{\bf v}\times {\bf E} \right].
\end{align}
For the sake of generality we will assume an infinite system with homogeneous $\epsilon>0$ and $\mu>0$. We then introduce two new fields as 
\begin{align}
{\bf F}^{(\pm)} = \sqrt{\epsilon}{\bf E} \pm i \sqrt{\mu}{\bf H},
\end{align}
the $\pm$ sign corresponds to photon helicity.
For time-independent velocity ${\bf v}$, we then get the Maxwells equations as 
\begin{align}
&\left({\bm \nabla} - \frac{\epsilon\mu - 1}{c^2}{\bf v}\partial_{t}\right) \times {\bf F}^{(\pm)} = \pm i\frac{\sqrt{\epsilon\mu}}{c}\partial_{t}{\bf F}^{(\pm)}, 
\label{MErot}
\\
&
{\bm \nabla}\cdot \left[ {\bf F}^{(\pm)} \pm \frac{\epsilon\mu - 1}{i\sqrt{\epsilon \mu} c} [{\bf v}\times {\bf F}^{\pm} ] \right] = 0 ,
\label{MEdiv}
\end{align}
which bear a similarity with Dirac equation, for a review of such approach see \cite{Iwo}. Here velocity ${\bf v}$ plays a role of a vector potential of an effective magnetic field. We assume velocity to have a cylindrical symmetry, described by 
\begin{align}\label{field}
{\bf v} = v(-y{\bf e}_{x}+x{\bf e}_{y}),
\end{align} 
where $v$ is angular velocity. Compare vector field (\ref{field}) describing rotation with the symmetric gauge of the magnetic field. 
As an example, one can keep in mind a dielctric medium of cylindric form, which is rotating about its axis. 
However, as mentioned above, we are going to study a rotating system infinite in all directions. 
Even though it is not experimentally possible, we wish to study this case in order to understand the nature of solutions. 
For finite systems it is straightforward to set boundary conditions by integrating Eq. (\ref{MErot}) and  Eq. (\ref{MEdiv}). 
We search for solutions in the form $\propto e^{-i\omega t}e^{ip_{z}z} $. 
For the sake of simplicity, introduce $\Omega = \frac{\sqrt{\epsilon \mu}}{c}\omega$ and $V = \frac{\epsilon \mu - 1}{c^2}v \omega$. 
In the following we choose $V  > 0$, and as mentioned above assume a system to be infinite in all directions. 
Components of equation (\ref{MErot}) are written as
\begin{align}
&
\Pi_{y} F_{z}^{(\pm)} - ip_{z}F_{y}^{(\pm)} = \pm \Omega  F_{x}^{(\pm)},
\\
&
ip_{z} F_{x}^{(\pm)} - \Pi_{x} F_{z}^{(\pm)} = \pm \Omega F_{y}^{(\pm)},
\\
&
\Pi_{x}F_{y}^{(\pm)} - \Pi_{y} F_{x}^{(\pm)} = \pm \Omega F_{z}^{(\pm)},
\end{align}
where $\Pi_{y} \equiv \left( -i\nabla_{y} +  Vx \right)$ and $\Pi_{x} \equiv \left( -i\nabla_{x} -  Vy \right)$ is the updated momentum operator. 
After straightforward transformations, assuming all components of ${\bf F}^{(\pm)}$ are non-zero, we obtain for $F_{z}^{(\pm)}$ component an equation
\begin{align} \label{schroedinger}
\left( \Pi_{y}^2  + \Pi_{x}^2 \right) F^{(\pm)}_{z} 
=\left( \Omega^2 -p_{z}^2 \right) F^{(\pm)}_{z}.
\end{align}
The equation has exactly the form of Schr\"{o}dinger equation describing an electorn in an uniform magnetic field, chosen to be described in a symmetric gauge \cite{LL3}. 
Hence we obtain the Landau solutions to the equation.
We label the eigen values and energies by index $n$, and write
\begin{align}
&
F^{(\pm)}_{z,n}  = e^{-V \vert \zeta \vert^2/2} \left( \partial_{{\bar \zeta}} - \frac{V}{2} \zeta  \right)^{n}f({\bar \zeta}),
\label{solutionF}
\\
&
\Omega_{n}^2 = 4V \left(n + \frac{1}{2}\right) + p_{z}^2 ,
\label{solutionO}
\end{align}
where $\zeta = x+ iy$, ${\bar \zeta} = x-iy$, and $f({\bar \zeta})$ is an arbitrary function of ${\bar \zeta}$. 
The function can be presented through basis states as
\begin{align}
f({\bar \zeta}) = \sum_{m} f_{m}({\bar \zeta}) = \sum_{m} \sqrt{N_{m}} {\bar \zeta}^{m},
\end{align}
where $N_{m}$ is a renormalization constant.
Each $f_{m}$ corresponds to $m$th orbit of the state on the $n$th level.
For example, for $n=0$ each $f_{m}$ corresponds to an orbit with a radius $\vert  \zeta \vert_{m} =\sqrt{ m/V}$.
 
Other two components of ${\bf F}^{\pm}$ are expressed through $ F_{z}^{(\pm)}$ as 
\begin{align}\label{xy1}
&
F_{x}^{(\pm)} =  \frac{\pm \Omega}{\Omega^2 - p_{z}^2}
 \left( i \Pi_{y} \pm \frac{p_{z}}{\Omega} \Pi_{x} \right) F_{z}^{(\pm)},
\\
&
\label{xy2}
F_{y}^{(\pm)} =  \frac{\pm \Omega}{i\left(\Omega^2 - p_{z}^2 \right)} 
\left(\Pi_{x}  \pm i\frac{p_{z}}{\Omega} \Pi_{y}  \right) F_{z}^{(\pm)}.
\end{align} 
The solutions found from equations (\ref{MErot}) are consistent with equations (\ref{MEdiv}). 
This can be seen by taking the divergence operation of the expression (\ref{MErot}).

\begin{figure}
 \centerline{\includegraphics[clip,width=0.8\columnwidth,angle=270]{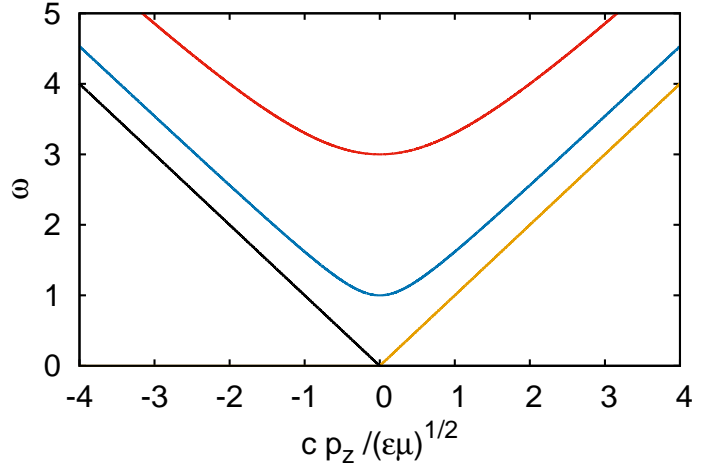}}
\protect\caption{ The spectrum of the Landau levels for electromagnetic waves. Levels $\omega_{n}$ described by \ref{omega} are plot in blue for $n=0$  and in red for $n=1$. Two branches of the lowest Landau level described by Eq. (\ref{helical}) are plot in black and yellow. This level is gapples and helical, i.e. black corresponds to $+$ helicity while yellow corresponds to $-$ velocity for $v>0$ choice of angular velocity of rotation. Parameter $2\frac{\epsilon \mu - 1}{\epsilon \mu} v = 1$ in appropriate units of frequency is chosen for sake of simplicity. }

\label{figure}

\end{figure}

Obtained spectrum, keeping in mind that $\omega>0$ and $\epsilon \mu - 1 > 0$, is rewritten in a more transparent form
\begin{align}\label{omega}
\omega_{n} 
&= 2\frac{\epsilon \mu - 1}{\epsilon \mu} v \left( n+\frac{1}{2} \right) 
\\
&
+ \sqrt{ \left(2 \frac{\epsilon \mu - 1}{\epsilon \mu} v \right)^2 \left( n+\frac{1}{2} \right)^2 + \frac{c^2p_{z}^2}{\epsilon\mu}}.
\nonumber
\end{align} 
Note that the spectrum does not depend on the helicity index $(\pm)$. 
Hence the obtained solutions are degenerate in helicity. 
See Fig. [\ref{figure}] for schematical description of the Landau levels for $n=0$ and $n=1$.

\textit{\underline{Helical mode}}.
In the previous subsection we assumed that all components of ${\bf F}^{(\pm)}$ are non-zero. 
We observe that there is an ambiguity of the expressions (\ref{xy1}) and (\ref{xy2}) if one sets $F_{z}^{(\pm)} = 0$ and $\Omega^2 = p_{z}^2$ in them.
Hence, in the following we examine the $F_{z}^{(\pm)} = 0$ case.  
In the following we will again consider inifinite geometry and assume propagation in $z-$ direction, such that ${\bf F}^{(\pm)} \propto e^{-i\omega t}e^{ip_{z}z}$.
 In this case we obtain for the equation (\ref{MErot}),
\begin{align}
&
 ip_{z}F_{y}^{(\pm)} = \mp \Omega F_{x}^{(\pm)},
\\
&
ip_{z} F_{x}^{(\pm)}  = \pm \Omega F_{y}^{(\pm)},
\\
&
\Pi_{x}F_{y}^{(\pm)} - \Pi_{y} F_{x}^{(\pm)} = 0,
\end{align}
from where one gets $F_{x}^{(\pm)} = is F_{y}^{(\pm)}$, with solution index $s=\pm$ which does not correspond to helicity. 
For $s=+$ we get $(\Pi_{x} - i\Pi_{y})F_{y}^{(\pm)} = -i(2\partial_{\zeta} + V{\bar \zeta})F_{y}^{(\pm)} = 0$, which can be met by $F_{y}^{(+)} = f({\bar \zeta}) \exp(-\frac{V}{2}\vert \zeta \vert^2)$ solution, with arbitrary function $f({\bar \zeta})$, given a requirment for the solution to decay as $\vert \zeta \vert \rightarrow \infty$.  
 For $s=-$ we get $(\Pi_{x} + i\Pi_{y})F_{y}^{(\pm)} = -i(2\partial_{{\bar \zeta}} - V\zeta )F_{y}^{(\pm)} = 0 $, which can not be met by solutions decaying at large $\vert \zeta \vert$. Hence, only the $s=+$ solution exists for $v>0$ choice of the angular velocity of dielectric media rotation.
The spectrum of $s=+$ solution for $\pm$ helicity is
\begin{align}\label{helical} 
\omega_{\pm} = \mp \sqrt{ \frac{c^2}{\epsilon\mu}} p_{z},
\end{align}
together with the $\omega > 0$ condition we obtain a requirement that $+$ helical photons must have $p_{z}< 0$, and that $-$ helical photons must have $p_{z}>0$. Hence, such solution is purely helical, i.e. opposite helicities counter-propagate. See Fig. [\ref{figure}] for schematics. If the rotation is switched to an opposite, $v\rightarrow -v$, the propogation structure of different helicities switch places. Note that the spectrum (\ref{helical}) satisfies the $\Omega^2 = p_{z}^2$ assumption which we started this subsection with. Hence solutions described by  $F_{z}^{(\pm)} \neq 0$ obtained in (\ref{solutionF}) and helical $F_{z}^{(\pm)} = 0$ solution are different.

\textit{\underline{Helical vortical effect for photons}}. 
Here we calculate equilibrium helicity current in the direction of the rotation axis. 
For that we assume a gas of photons propagating inside hypothetical infinite and rotating dielectric medium.
The current of a given photon helicity is 
\begin{align}
j_{z}^{[\pm]} = \sum_{n} \int_{-\infty}^{+\infty} \frac{dp_{z}}{4\pi^2} \left( \frac{\epsilon\mu - 1}{c^2} \vert v\vert \omega_{n}  \right) \frac{d \omega_{n}}{dp_{z}} g(\omega_{n}),
\end{align}
where a $\left( \frac{\epsilon\mu - 1}{c^2} \vert v\vert \omega_{n}  \right)>0$ factor is due to the Landau level degeneracy \cite{LL3}, $\frac{d \omega_{n}}{dp_{z}}$ is the photon velocity along $z-$ direction, and $g(\omega_{n})= \left( e^{\omega_{n}/T} -1 \right)^{-1}$ is the Bose-Einstein distribution function at non-zero temperature $T$. 
Summation is over all Landau levels obtained in previous sections and given by Eqs. (\ref{omega}) and (\ref{helical}).
Only the lowest Landau level given by Eq. (\ref{helical}) is helical, hence it is the only level that contributes to the helicity current. 
Calculations show that the helicity current is
\begin{align}
j^{[\mathrm{H}]}_{z} = j_{z}^{[-]} - j_{z}^{[+]} = \left( \frac{\epsilon\mu - 1}{c^2} v \right)\frac{T^2}{12} .
\end{align}
This current means that in hypothetical infinite and rotating dielectric medium opposite helicities of photons will counter-propagate. Net photon current is zero, i.e. $ j_{z}^{-} + j_{z}^{+} =0 $ as epxected in equilibrium.

\textit{\underline{Discussion}}. 
For inhomogenous case when $\epsilon$ and $\mu$ are functions of coordinates, the Eq. (\ref{MErot}) and Eq. (\ref{MEdiv}) will change, see for review \cite{Iwo}. For example imagine a rotating cylinder, in which case outside of the cylinder $\epsilon = 1$ and $\mu=1$, and inside $\epsilon \neq 1$. Helicities will then be mixed due to the inmhomogeneous $\epsilon$ and $\mu$ functions. Helicity degeneracy of solutions of Maxwell equations described by (\ref{solutionF}) and (\ref{omega}) will not change in finite geometry. However, helical mode solutions (\ref{helical}) will change due to helicity mixing. Due to boundary conditions on the walls of rotating cylinder, it might be experimentally hard to observe the helical mode solutions. Also, it is important to note that there is a natural limit on the radius of rotating cylinder given by condition $\vert {\bf v} \vert/c \ll 1$. Therefore, if it will be possible to excite such helical mode in finite geometry, there will be spacial separation of co-propagating opposite helicity waves. For example in particular direction of cylinder rotation, intensity of $+$ helicity of the $p_{z}>0$ wave will be peaked at the surface of the cylinder, while intensity of $-$ helicity will be peaked closer to the axis of the cylinder.   

We note a striking similarity of the obtained in the present paper helical mode to the chiral lowest Landau level of three dimensional Dirac fermion, for example see \cite{NN}. The similarity is due to non-trivial Berry curvature of photons and Dirac fermions. In a rotating gas of photons opposite helicities counter-propagate along the axis of rotation. It is the helical vortical effect, an analog of chiral magnetic effect for Dirac fermions, see \cite{Kharzeev} for a review.

From the structure of the Maxwell equation (\ref{schroedinger}) we observe that effective charge of photon is its frequency, while the rotation plays a role of effective vector potential for the photon. Obtained conditions $(\Pi_{x} - is\Pi_{y})F_{y}^{(\pm)} = 0$, with $s=\pm$ denoting relation $F_{x}^{(\pm)} = is F_{y}^{(\pm)}$, for the helical mode are helicity sensitive due to rotation. It can be checked that the solution becomes helicity degenerate when the rotation is switched off, as is expected for free space propagating photon.    

Recently there were experiments on creating synthetic Landau levels for photons using resonators \cite{synthetic1}. Various proposals on creating the angular momentum for photon \cite{Stone} and \cite{synthetic2} were put forward.   The author believes that the obtained in the present paper results will excite further experimental interest in studies of Landau levels for photons.

It is tempting to search for photon Landau levels in pulsars, however we note the dielectric function in the pulsar atmosphere is $\epsilon \sim 1$, and the levels vanish due to $\epsilon\mu - 1$ factor.  It is possible that Landau levels for photons proposed in this paper can be observed in experiments with slow light \cite{slowlight}. It requires further thorough investigation.

\underline{\textit{Acknowledgements}}.
The author is greatful to P.S. Shternin and D.G. Yakovlev for discussions on astrophysics, and to A.Yu. Zyuzin for discussions on slow light.
VAZ acknowledges A.F. Ioffe Institute for warm hospitality during winter.

\underline{\textit{Note added}}.
Calculation of helical vortical effect (chiral vortical effect) for photons appeared in the second version of this paper, shortly after papers \cite{AvkhadievSadofyev,Yamamoto}.
VAZ believes the results of \cite{AvkhadievSadofyev,Yamamoto} and of the present paper, all being obtained by different methods, compliment each other. 
Present paper utilizes the zero mode description of helical vortical effect.

\end{document}